\documentclass[10pt,letterpaper,twocolumn]{revtex4} 

\usepackage{graphicx}
\usepackage{dcolumn}
\usepackage{multirow}
\usepackage{bm}
\usepackage{mathrsfs}

\begin{document}


\title{Imaging using nano metallic films: from evanescent wave lens to resonant tunnelling lens}

\author{Jensen Li, C.T. Chan}

\affiliation{Department of Physics, The Hong Kong University of Science and Technology, \\
Clear Water Bay, Kowloon, Hong Kong, China}


\begin{abstract}
A thin metal slab is known to be able to focus the near fields of TM wave of a point source. 
Here, we show that a thin metal slab in fact possesses a far-field image and through a simple 
modification on the system by resonance tunneling, a double metal slab can give a bright image 
with both the far-field and the near-field details.
\end{abstract} 



\maketitle

\noindent 
Double-negative (DNG) medium, with permittivity ($\epsilon$) and permeability 
($\mu$) simultaneously negative, exhibits a negative value of refractive 
index \cite{Veselago:1968,Pendry:2000} and the resulting negative refraction 
has been demonstrated both numerically \cite{Ziolkowski:2001} and experimentally 
\cite{Shelby:2001,Parazzoli:2003}. 
The DNG slab with $\epsilon = \mu = - 1$ is a perfect lens \cite{Pendry:2000} 
in the sense that it can focus all 
geometric beams (the far fields) of different directions from a point 
source to a real image by negative refraction and it can 
compensate exactly all the evanescent waves (the near fields). 
The experimental realization of the perfect lens is difficult due to its extreme 
sensitivity to the system parameters \cite{Smith:2003}. 
On the other hand, if we are working in the near-field regime 
for TM waves, the condition $\epsilon = - 1$ is sufficient so that a 
thin metallic film acts as a near-field lens \cite{Pendry:2000,Fang:2005,Blaikie:2006}. 
However, for the metallic film, the position of the focal plane for the far fields is ambiguous 
although it is a common practice \cite{Fang:2005,Shen:2002} to place the image 
plane at a distance double the slab thickness beyond the point source. 
The position of the focal plane is proposed to be determined 
from the phase contour or the streamline of Poynting vector 
\cite{Ziolkowski:2001,Shamonina:2001} but whether a far-field image 
exists remains unclear.

\begin{figure}[htbp]
\centerline{\includegraphics[width=2.8in]{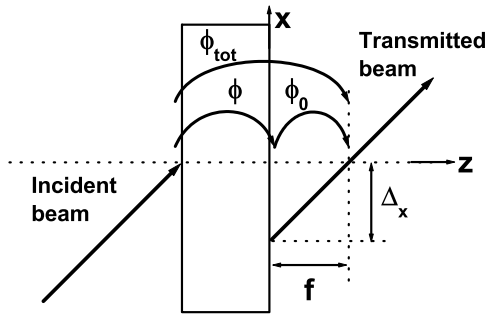}}
\caption{A transmitted beam undergoes negative refraction for a slab of thickness d. 
$\Delta _x $ is the lateral beam shift. 
$\phi $ is the phase change across the slab. $\phi _0 $ is the phase 
change in vacuum and $\phi _{tot} $ is the total phase change for the beam in returning to the z axis.
}
\label{fig1}
\end{figure}

In this paper, we take an alternative approach by considering the far-field focusing 
of a thin slab with arbitrary $\epsilon$ and $\mu$. We show that if the slab 
can be regarded as a lens, the phase of the source-to-image transfer function can only be certain quantized values. 
The $\epsilon = \mu = -1$ perfect lens is a special case of the general condition. 
It follows naturally from our results that a ``single negative'' metallic film can 
focus a point source to a far-field image. 
In addition, a Fabry-Perot-like cavity of two metallic films can provide both 
far-field and near-field imaging with high transmittance. 
Here, a far-field image means the image formed by all the propagating 
waves and it can be recorded (e.g. by a camera) far away from the lens.

Let us first establish the condition for negative refraction (a negative 
lateral shift) for a flat slab of pre-specified values of 
permittivity $\epsilon $ and permeability $\mu $. 
Without loss of generality, we consider TM waves. 
Fig. \ref{fig1} shows a schematic diagram showing the situation when a geometric beam 
impinges on this slab.
$f$ here is defined as the distance measured from the exit interface of the 
slab when the beam returns to the z-axis and 
$\Delta _x $ is the lateral beam shift.
For negative (positive) refraction, $f$ is positive (negative) and 
$\Delta_x$ is negative (positive). 
Using the stationary-phase method \cite{Goos:1947}, $f$ can be proved to be given by

\begin{equation}
\label{eq1}
f\left( {k_{0z} } \right) = - \frac{\partial \phi }{\partial k_{0z} },
\end{equation}

\noindent
where $\phi $ is the phase change across the slab and we use $k_z $($k_{0z})$ 
to denote the wave number in the z-direction in the slab (vacuum). In 
general, $f$ is a function of the incident angle where $k_{0z} $ 
ranges from 0 at grazing incidence to $k_0$ (the wave number in vacuum) at 
normal incidence.

\begin{figure}[htbp]
\centerline{\includegraphics[width=3.9in]{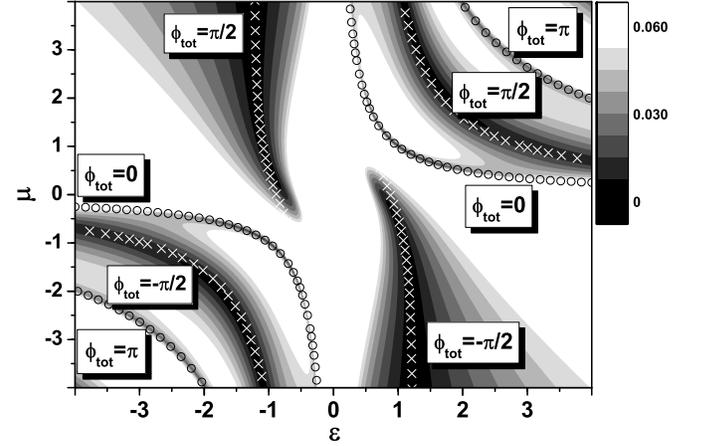}}
\caption{Lensing condition for a slab of thickness $d = 0.19\lambda $. 
The open circles and crosses show the analytic results. 
The grayscale shows the operational aberration ($a / \lambda$)
defined in text with black color denoting smaller values of aberration.}
\label{fig3}
\end{figure}

Now, we search for optimal configurations such that $f$ is nearly independent 
of the incident angle (measured by $k_{0z}$). Such an optimal configuration 
implies the slab behaves like a lens, and the lensing condition can be restated in 
terms of the phase change as

\begin{equation}
\label{eq3}
\frac{\partial \phi _{tot} \left( {k_{0z} } \right)}{\partial k_{0z} } = 
k_{0z} \frac{\partial f\left( {k_{0z} } \right)}{\partial k_{0z} } \approx 
0\mbox{,}
\end{equation}

\noindent
where $\phi _{tot} $ is the total phase change when the beam returns to the z-axis.
We can measure how well Eq. (\ref{eq3}) can be satisfied by defining an operational aberration using

\begin{equation}
\label{eq7}
a = \sqrt {\frac{1}{k_{0}^3}\int_0^{k_0 } {\left( {f\left( {k_{0z} } \right) 
- f(k_0)} \right)^2 3 k_{0z}^2 d} k_{0z} } ,
\end{equation}

\noindent
which measures the deviation of $f(k_{0z})$ from its para-axial value. 
For a slab of thickness $d = 0.19 \lambda$, the values of $a$ are plotted in grayscale in Fig. \ref{fig3} 
with black color denoting smaller values. The loci of the local minima 
of $a$ in the $\epsilon - \mu $ plane defines the configurations for 
good focusing. Interestingly, they form bands and each band can be characterized according to the 
total elapsed phase by

\begin{equation}
\label{eq4}
\phi _{tot} \left( {k_{0z} } \right) \approx m\pi / 2,
\end{equation}

\noindent
where $m$ is any integer and the approximation is valid for all $k_{0z}$ ranging 
from $0$ to $k_0$. The stationary $\phi_{tot}$ is thus quantized. 
In fact, the bands can be approximated by analytic formulas. We will not prove them in details but simply 
state them here. We found that there are two scenarios that can lead to 
Eq. (\ref{eq4}). The first scenario is given by

\begin{equation}
\label{eq5}
\chi = m \pi,
\end{equation}

\noindent
where $\chi$ is defined by $\chi = \sqrt {\epsilon \mu - 1} k_0 d$ 
and $m$ is an integer. Eq. (\ref{eq5}) yields the bands of 
$\phi_{tot} = m \pi$ in the $\epsilon$-$\mu$ plane.
These bands for a slab of $d = 0.19\lambda$ are plotted as black open 
circles in Fig. \ref{fig3}. 
The two bands of $m=0$ correspond to a transparent slab of refractive 
index +1 and a DNG slab of refractive index --1. 
The second scenario is given by

\begin{equation}
\label{eq6}
\epsilon ^2 = - \frac{3\tan \left( \chi \right)}{2}\frac{2\chi + \sin 
\left( {2\chi } \right)}{7 + \cos \left( {2\chi } \right)}, 
\end{equation}

\noindent
which yields the bands of $\phi_{tot} = (m+1/2)\pi$. These bands are plotted 
as white crosses in Fig. \ref{fig3}.
In Eq. (\ref{eq6}), $\chi$ can be either purely real or purely imaginary from its definition.

We see that the bands of $\phi_{tot} = (m+1/2) \pi$ is in general broader 
than the bands of $\phi_{tot} = m \pi$ (which includes $n = -1$ as a special case). 
The DNG lens with $\epsilon = \mu = -1$ is the ``best lens'' since it has 
exactly zero aberration, but it is also the least robust. The ``poor man's superlens'' 
show up as a special case as the band $\phi_{tot} = \pi / 2$. Only 
$\epsilon$ is negative (single-negative), but the aberration can be very small. 
In particular, we select the case $\left( {\epsilon ,\mu } \right) = \left( { - 1.0914,1} 
\right)$ along this band. It has the smallest aberration of $a = 2\times 10^{ - 4}\lambda$ with 
$f = 0.256\lambda $.
The lens formula, like the DNG lens, is governed by $s_1 + s_2 = f$ 
where $s_1$ ($s_2$) is the distance of the point source (image) from the first 
(second) interface. $s_2 $ is positive (negative) for a real (virtual) image.
If we denote $t_{tot}$ as the complex transfer amplitude from the source to the image for the lens, 
the image is formed due to the phase conservation of $t_{tot}$ at different 
incident angles and it does not require the amplitude 
of $t_{tot}$ to be a constant. 
In the current example, the amplitude varies from 0.3 at normal incidence to 0 at grazing incidence. 
As the image is formed by tunnelling, $f$ stays nearly the same 
as we increases the slab thickness (unlike the DNG slab where $f$ scales with the slab thickness).
However, the transfer amplitude drops exponentially with the slab thickness.

\begin{figure}[htbp]
\centerline{\includegraphics[width=3.3in]{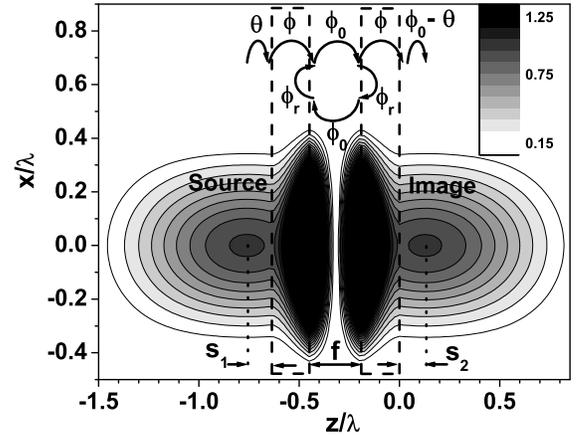}}
\caption{The contour of the far-field E-field profile (square magnitude) at 
resonance-tunnelling focusing. Black (while) color denotes the high (low) values. 
The curved arrows show the phase change of the wave in different regions.}
\label{fig4}
\end{figure}

\begin{figure}[htbp]
\centerline{\includegraphics[width=3.5in]{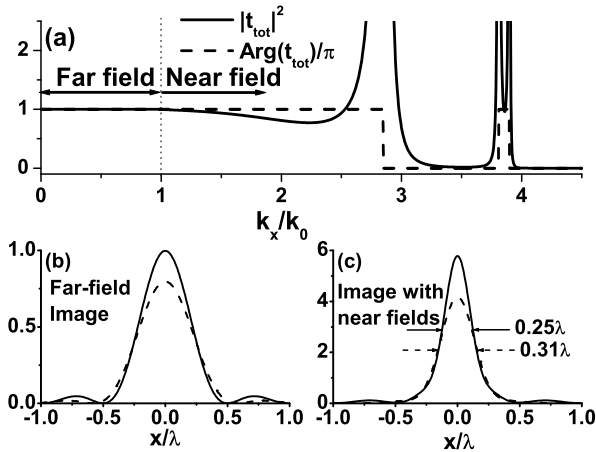}}
\caption{(a) The source-to-image transfer function. Solid line 
shows the square magnitude and dashed line shows the phase; 
(b) and (c) show the image profiles at the image plane with far 
fields only and with near fields included in solid lines. Dashed 
lines show the profile (4x magnified) when 0.1i is added to the 
permittivity of the metal.
}
\label{fig5}
\end{figure}

We have shown that a metal slab can have a far-field image. 
Moreover, an example of optimal aberration has been given explicitly. Based on the optimal 
example, the brightness of the image can in fact be increased to nearly unity using an interesting 
property of tunnelling focusing. 
As the slab has a mirror symmetry in the z-axis, the reflection phase $\phi_r$ from 
the slab entrance interface satisfies $\phi _r = \phi \pm \pi / 2$. 
If we stack another slab of the same configuration at precisely one $f$ 
beyond the original slab (details shown in Fig. \ref{fig4}), the phase change per 
cycle within the cavity is $2(\phi_r+\phi_0) = 0$ or $2\pi$. 
It is a resonating condition satisfied for all incident angles, resulting 
in an all-angle (or equivalently a real-space) Fabry-Perot (FP) cavity. 
Moreover, a real image is formed at one $f$ beyond the second slab 
with $\phi _{tot} = 2\left( {\phi + \phi _0 } \right) = \pi$ 
which is independent of the incident angle.
As an example, if we stack two identical metal slabs with $\epsilon = -1.0914$ 
and thickness $d = 0.19\lambda $ together with an air gap equal to one $f$ 
of that a single slab in between, 
we can form a real image with $s_1 + s_2 = 0.256\lambda$. 
Fig. \ref{fig4} illustrates the source and the image profile for this double-slab
system if we place a point source at $f/2$ in front of the first 
slab. Only propagating fields from the point source are included in our 
calculation. The intensity at the source point is normalized to one and 
nearly the same intensity is found at the image. 
In principle, the propagating wave transmittance is essentially unity 
independent of the incident angle for thin slabs, but if thicker metal 
slabs are employed, the transmittance of the double-slab system becomes more sensitive to the 
aberration the single-slab system.

We now consider the evanescent wave components.
Fig. \ref{fig5}(a) shows $t_{tot}$ of the same system plotted 
against $k_x$, the wave number in the transverse direction. 
From $k_x=0$ to $k_0$ (the propagating waves), $t_{tot} \approx -1$ 
for all incident angles. Therefore, the image profile 
for a far-field observer is simply $(sin(k_0 x)/(k_0 x))^2$ 
as shown by the solid line in Fig. \ref{fig5}(b). Therefore, for a far-field observer, 
the double-slab system behaves just like the DNG perfect lens. For the near 
fields with $k_x >1$, the double-slab system can also support surface plasmons, 
indicated by the peaks shown in Fig. \ref{fig5}(a), and this leads to 
super-resolution. The image profile with the evanescent waves 
included is shown by the solid line in 
Fig. \ref{fig5}(c). It has a half-pitch resolution of $0.25 \lambda$ 
which breaks the diffraction limit.
The dashed lines in Fig. \ref{fig5}(b) and (c) show the image profile 
if we introduce some absorption (an extra $0.1i$ is added to the 
permittivity of the metal) into the system. The lensing effect is still retained.
A double-layer structure is currently employed for reducing the 
absorption in near-field focusing in the literature \cite{Blaikie:2006} 
Our results show that a similar system can work for the far fields as 
well by tuning it into FP resonance. The image can be transferred 
farther away by cascading a number of these cavities.

In summary, we have considered the general lensing condition for a homogeneous 
slab, which is found to be discrete values of the transmission phase. 
For a metal slab capable in focusing, an all-angle FP cavity can 
be constructed by stacking two of them.
A far-field image with high transmittance is formed. 
The double-slab system is an improved version of the near-field lens from 
a single metallic film as far-field focusing can also be obtained.

\section*{Acknowledgement}

This work is supported by RGC Hong Kong through CA02/03.SC01. 
Jensen Li is also supported by the Croucher Foundation fellowship from Hong Kong.

\end{document}